\documentclass[journal,12pt,onecolumn,draftclsnofoot]{IEEEtran}
\usepackage{graphicx}
\usepackage{amsmath}
\usepackage[linesnumbered,ruled]{algorithm2e}
\usepackage{amsfonts}
\usepackage{amssymb}
\usepackage{bm}
\usepackage{cite}
\usepackage{color}
\usepackage{multirow}
\usepackage{tabularx}
\usepackage{comment}
\usepackage{mathtools}

\newtheorem{prob}{Problem}
\newcolumntype{L}[1]{>{\raggedright\arraybackslash}p{#1}}
\newcolumntype{C}[1]{>{\centering\arraybackslash}p{#1}}
\newcolumntype{R}[1]{>{\raggedleft\arraybackslash}p{#1}}

\usepackage{times, epsfig}
\usepackage{subfig}
\newlength{\figwidth}
\newtheorem{lemma}{Lemma}

\begin{document}
%\baselineskip 20 pt

%\title{Optimal Energy Harvesting and Power Allocation Policy for Wireless Powered Sensor: A Perspective from Age of Information}

\title{Minimum Age of Information in Internet of Things with Opportunistic Channel Access}
%Scheduling for optimal Age of Information in Wireless Opportunistic Channel Access Networks
%elements in title: wireless sensors networks, mutiple nodes, opportunistic channel acess, Age of Information, optimal Scheduling policy.
\author{Lei Wang and Rongfei Fan

%\vspace{-10mm}

\thanks{

%XX
L. Wang, is with School of Information and Electronics, Beijing Institute of Technology, Beijing 100081, P. R. China. (wanglei1995bit@gmail.com)

R. Fan, is with School of Cyberspace Science and Technology, Beijing Institute of Technology, Beijing 100081, P. R. China. (fanrongfei@bit.edu.cn).
}
}

%\date{}
\maketitle

%%%%%%%%%%%%%%% Article Body %%%%%%%%%%%%%%%%%%%%%%%%%%%%%%%%%%%%%%%%%

%%%%=======================================================

\begin{abstract}
This paper investigates an Internet of Things (IoT) system in which multiple devices are observing some object's physical parameters and then offloading their observations back to the BS in time with opportunistic channel access. Specifically, each device accesses the common channel through contention with a certain probability firstly and then the winner evaluates the instant channel condition and decides to accept the right of channel access or not. We analyze this system through the perspective of Age of Information (AoI), which describes the freshness of observed information. The target is to minimize average AoI by optimizing the probability of device participation in contention and the transmission rate threshold. The problem is hard to solve since the AoI expression in fractional form is complex. We first decompose the original problem into two single-variable optimization sub-problems through Dinkelbach method and Block Coordinate Descent (BCD) method. And then we transform them to Monotonic optimization problems by proving the monotonicity of the objective functions, whose global optimal solution is able to be found through Polyblock algorithm. Numerical results verify the validity of our proposed method.
\end{abstract}


%\vspace{-4mm}
\section{Introduction} \label{s:intro}
In recent years, with the development of the Internet of Things, more and more intelligent systems, e.g., smart home, connected vehicle, smart industry, etc., have been focused on.
%In order to facilitate various intelligent IoT applications, wireless sensors are deployed to observe and monitor the status of environment or equipment, such as temperature, humidity, speed, etc., and then upload the information to a base station \cite{IoT}.
In order to facilitate various intelligent IoT applications, some remote devices are deployed to observe and monitor the status of environment or equipment, such as temperature, humidity, speed, etc., and then upload the information over a wireless network \cite{IoT}.

In many real intelligent systems, such as connected vehicle, smart industry, they are very sensitive to latency.
But some traditional metrics, e.g., delay, throughput, are no longer sufficient for the time-sensitive characterization of these applications. To capture the information freshness, a new metric, Age of Information (AoI), has received much attention recently \cite{Trends}.
AoI represents the elapsed since the generation of an observation which was most recently received by a destination.
Its averaging over time is usually used to analyze and optimize system information freshness \cite{AoI1}.
%However, due to the different location and environment of each sensor in the IoT systems, the channels between the sensors and the base station are varied. Thus, in order to improve channel utilization and reduce latency, opportunistic channel access (a user with poor channel quality gives up its channel access opportunity to other users with good channel conditions) has been concerned in much literature because of its efficient utilization of channel resources \cite{opportunistic1} \cite{opportunistic2}.

%Meanwhile, some traditional metrics, e.g., delay, throughput, are no longer sufficient for the time-sensitive characterization of these applications. To capture the information freshness, a new metric, Age of Information (AoI), has received much attention recently \cite{Trends}.
%more proper representation
%AoI can be defined as $t-u(t)$, where $u(t)$ represents the generation time of the information at the source, which has reached its destination before time $t$. Its averaging over time is usually used to analyze and optimize system information freshness  \cite{Trends} \cite{AoI1}.

Early researches of AoI focused more on single device scenarios. However, as IoT applications become more widespread and networks become more complex, the allocation of channel resource for a large number of devices deserves to be researched. References \cite{multichannel, scheduling, aloha2, aloha3, aloha1} all investigate multiple access networks with the AoI performance.
%The works in \cite{FDMA} investigates the minimum average AoI in multiple devices networks based on FDMA. 
The work in \cite{multichannel} gives comparative conclusions on the advantages and disadvantages of TDMA and FDMA respectively and indicates that multichannel access can provide low average AoI and uniform bounded AoI simultaneously. 
%The work in \cite{scheduling} investigates the AoI performance in an orthogonal channels without collisions transmission IoT system with different scheduling policies.
The work in \cite{scheduling} investigates the AoI performance in an orthogonal channels without collisions transmission IoT system with three different scheduling policies, Greedy policy, Max-Ratio policy, and Lyapunov policy.
However, these multiple access protocols cannot meet the requirements of massive IoT networks due to signaling and control overhead which increases with network size. Therefore, many researches investigate the AoI performance with random access. %such as/is a protocol/直接ALOHA 
Slotted ALOHA, which is a widely utilized random access protocol because of its simplicity and low cost in implementation, allow users to share the channel without any coordination. Specifically, in slotted ALOHA, sources that wish to send data transmit with a certain probability in each slot.
%, such as ALOHA and Carrier-Sense Multiple Access (CSMA).
%In \cite{aloha1}, the average AoI in the slotted ALOHA protocol is derived. The collisions, defined as more than one device transmit, is characterized as interference into the SINR and the transmission success probability is derived by comparison of the SINR and its threshold.
%Also in slotted ALOHA, different from \cite{aloha1}, in \cite{aloha2}, once a collision occurs, all the transmission data are lost and the device have to try a retransmission.
In \cite{aloha2}, the average AoI in the slotted ALOHA protocol is derived, and it is assumed that once there is only one device transmitting in a time slot, the transmission must be successful and finish in one time slot.
The research \cite{aloha3} investigate an ALOHA networks with limited retransmissions. All links experience interference with the same distribution, so the transmission success probability is constant and the device need to retransmit its updates.
More specifically, \cite{aloha1} consider that the transmission success probability is depends on the SINR and a certain threshold, which is changed due to the interference. 
%The work \cite{csma} and \cite{csmaca} are both investigate the CSMA random access system.
%The work \cite{csma} investigate the optimal back-off time in CSMA networks to minimize the average AoI.
%And the work \cite{csmaca} analyze the worst-case AoI in a CSMA/CA network, a CSMA with collision avoidance system.
%Similar in work \cite{aloha3}, the letter analyze the mean peak AoI by optimizing retransmission rate and the number of allowed retransmissions.
%And in \cite{aloha4}, a slotted ALOHA system with Gilbert-Elliot channel model is analyzed on a power-law AoI perspective.
And all these works make an assumption that the period of per transmission is constant and not related to the channel state.

%Recently, AoI metrics have been widely deployed in systems analysis or resource allocation and scheduling in IoT systems. References \cite{multichannel, FDMA, nocontend, contend} all investigate resource allocation and %scheduling problems in IoT systems based on the AoI perspective.
%The works in \cite{multichannel} investigates a multi-user system on average AoI and bounded AoI perspective. It demonstrates that a multichannel system with the same received power has better performance in AoI and it gives comparative conclusions on the advantages and disadvantages of TDMA and FDMA respectively.
%The works in \cite{FDMA} investigates the channel resource allocation problem in multi-node constant channel state IoT systems based on AoI, where sensor nodes upload data in the way of FDMA.
%The works in \cite{nocontend} investigates the AoI problem of collision-free access for time-varying channel with multiple users based on three sampling and scheduling policies: Greedy, Max-Ratio and Lyapunov.
%The works in \cite{contend} investigates a periodic synchronous sampling AoI optimization problem under a multi-user system, where users access the channel through a random Aloha-like access method with collisions.
%However, AoI in multi-user asynchronous sampling IoT systems with opportunity channel access has not been concerned yet.

Different from these researches, this paper consider a opportunistic channel access, which is similar to the slotted ALOHA, IoT system, in which channel state is varying with time. 
%in which the channel state changes can affect the transmission period and the AoI performance.
Thus, it need a transmission threshold to determine whether waiting or transmitting in order to limit the time delay and optimize the AoI.
%Thus, different from \cite{multichannel, FDMA, nocontend, contend} above, this paper considers a multi-node non-synchronous system with opportunistic channel access and solves the problem of minimum AoI. In particular, different from \cite{contend}, this paper considers the impact of channel state on AoI.
%Specifically, we consider a system in which multiple wireless sensors observe the environment and transmit updates to the base station. Each sensor obtains the opportunity to access the channel through a random contention with certain probability firstly.
%After the winning sensor obtains the opportunity, the channel capacity between it and base station with the current channel state will be compared with a threshold to determine whether the update can be transmitted at this time.
In this framework, what probability that devices will attend contention with and how to set the transmission rate threshold are of importance for minimization of the AoI.
%This optimization problem is hard to solve because the expression of AoI is complicated since it involves many random variables.
We present a mathematical framework for analyzing the average AoI in this system and discuss the impact of the access probability and transmission threshold on the average AoI.
The challenge for the problem is to derive the closed form expression of the average AoI and give the optimal solution of this complicated formulation, which is overcome through Dinkelbach method, Block Coordinate Descent (BCD) method, and Monotonic optimization method.
%So we discuss the independence and distribution of random variables firstly, and then transform the objective function from fractional formulation to a non-fractional problem with the same solution by Dinkelbach's method. The optimal solution obtained by the monotonic optimization method is given along with the monotonicity analysis of the average AoI for contention probability and rate threshold respectively.

\section{System Model and Problem Formulation} \label{s:model}

Consider an IoT system with one base station (BS) and $N$ IoT devices, which constitute the set $\mathcal{N} \triangleq \{1, 2, ..., N\}$.
These $N$ IoT nodes have the obligation of observing some object's physical parameters and then offloading their observations back to the BS in time.
In terms of observation offloading, all the $N$ IoT nodes share a common channel with bandwidth $B$, to access into the BS.

\begin{figure}[!hbtp] 
	\begin{center}
		\includegraphics[angle=0,width=0.8 \textwidth]{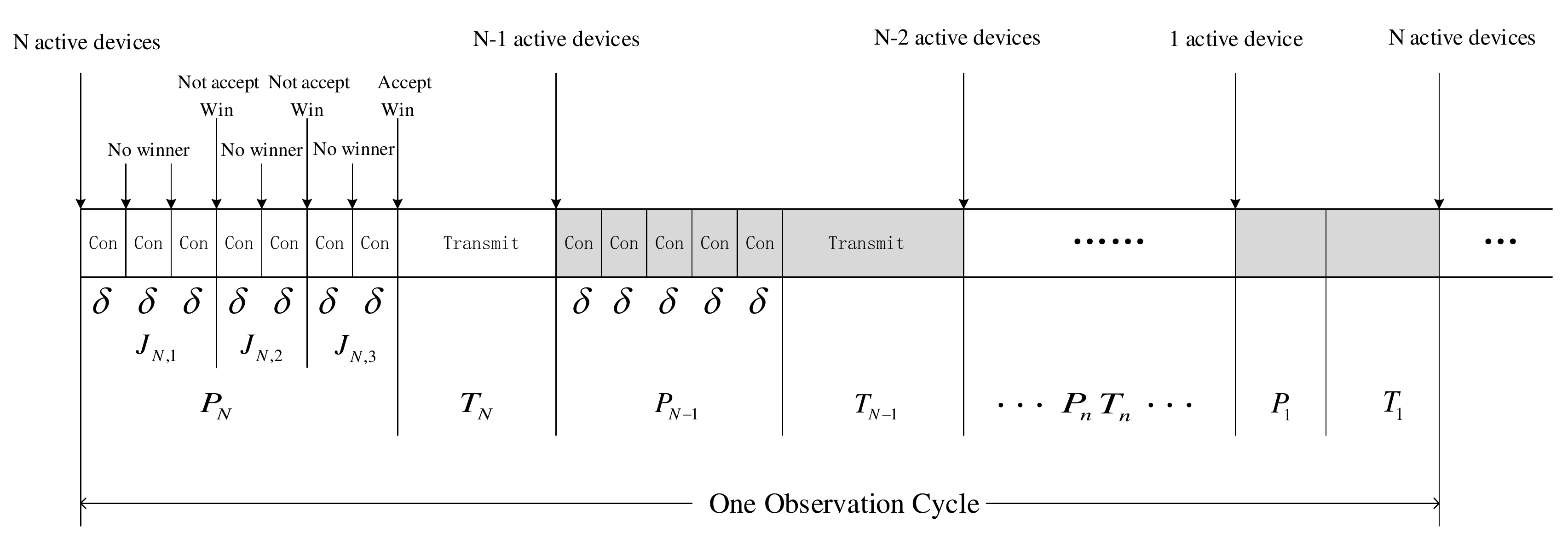}
	\end{center}
	\caption{Time slot of system.}
	\label{fig:timeslot}
\end{figure}

To be fair with every IoT device and keep observations fresh, the observing and data offloading is performed over multiple rounds in the following way, which is also illustrated in Fig. \ref{fig:timeslot}. In each round, there is only one IoT device who can win the right of exclusive use of the common channel and then perform observation and data offloading.
After offloading its observation, the winning IoT device will be inactive in subsequent rounds until all the rest IoT devices have offloaded their observations to the BS.
Through this operation, in each round there is one IoT device quits the observation and data offloading process.
We denote there rounds as Round $N$, $N-1$, ..., 1, respectively, and take these $N$ rounds as one observation cycle.
At the end of one observation cycle, all these $N$ IoT devices are inactive and another observation cycle will be initialized immediately.
%For this winning IoT device, it may select to offload its most updated observation in current round and then keep hibernated until all the rest IoT devices have offloaded their observations to the BS, it may also drop the chance in current round (if it is not satisfied with the channel quantity) and enroll the competition for the right of channel access again in the next round. Through this way, every IoT device has an opportunity to offload its observations timely.

In one of aforementioned round, say Round $n$, there are $n$ IoT devices having yet to offload their observations to the BS, i.e., $n$ active IoT devices, three steps of operations can be expected as follows:
\begin{itemize}
\item Step 1: These $n$ IoT devices compete for the right of exclusively accessing the common channel in an opportunistic way.
\item Step 2: The winning IoT device, denoted as $n'$th IoT device, evaluates the instant channel condition and then decides to accept the right of channel access or not.
\item Step 3: If the $n'$th IoT device accepts the right of channel access, it will make an instant observation on the interested object and offload it to the BS. If the $n'$th IoT device does not accept the channel access right, it will go back to Step 1 and start a new competition with the rest active IoT devices, which indicates that there may be multiple competitions of channel accessing right among these $n$ IoT devices in current round.
\end{itemize}

For the competition in Step 1, divide the time horizon into slots with equal length $\delta$. In every time slot, these $n$ IoT devices contend for the channel accessing right by broadcasting a pilot with probability $p$. Before the emergence of an unique winner, one of the following three possible cases may happen:
\begin{itemize}
	\item Case I: There is no IoT device broadcasting a pilot in current time slot, which happens with probability $(1-p)^n$. In such a case, these $n$ IoT devices will proceed to contend in next time slot.
	\item Case II: There are more than one IoT devices broadcasting a pilot in current time slot. In such a case, a collision happens and no IoT device wins, these $n$ IoT devices will proceed to contend in next time slot.
	\item Case III: There is only one IoT device broadcasting a pilot in current time slot, which happens with probability $p_n(p)=n p (1-p)^{n-1}$. In such a case, no collision happens and this unique IoT device wins the competition.
\end{itemize}

For the channel condition evaluation in Step 2, suppose the channel gain between each IoT device and the BS experiences independent and identical Rayleigh fading. Moreover, as assumed in \cite{optimalstopping}, the channel gain over disjoint time slots are also independently and identically distributed.
With these assumptions, denote $g_{n,i}$ as the instant channel gain between the winning IoT device of $i$th competition for the right of channel access in Round $n$ and the BS, then the distribution function of $g_{n,i}$ can be written as $f(g_{n,i}) = \lambda e ^{-\lambda g_{n,i}} \cdot\bm{1}(g_{n,i}\geq 0)$ where $\bm{1}(\cdot)$ is the indicator function.
Denote $p_T$ as the transmit power of every IoT device and $\sigma^2$ as the power spectrum density of noise, a transmit rate
\begin{equation}
R_{n,i} = B \ln\left( 1 + \frac{p_T g_{n,i}}{B \sigma^2}\right)
\end{equation}
can be realized between the winning IoT device and the BS.
In order to limit time delay of data offloading, a threshold of $r$ is imposed on transmit rate, i.e., the winning IoT device will drop the right of channel access if $R_{n,i}<r$. With such a setup, the probability of accepting the channel accessing right, denoted as $q(r)$, can be written as
\begin{equation}
	\begin{aligned}
		q(r) = \Pr \left( R_{n,i} \geq r \right) =\Pr \left( g_{n,i} \geq  {\left(e^{\frac{r}{B}} - 1 \right) B \sigma^2 }/{p_T} \right)
		=\int_ {{\left(e^{\frac{r}{B}} - 1 \right) B \sigma^2 }/{p_T}}^\infty  f(g_{n,i})d{g_{n,i}} =e^{ - \frac{\lambda {\left({e^{\frac{r}{B}}} - 1 \right) B{\sigma ^2}}}{p_T}}
	\end{aligned}
\end{equation}
which is a decreasing function with $r$.
If the winning IoT device accepts the channel accessing right, which is supposed to happen in $K_n$th competition, $g_{n, K_n}$ should be larger than $G(r) \triangleq {\left(e^{\frac{r}{B}} - 1 \right) B \sigma^2 }/{p_T}$, the distribution function of channel gain in this case, denoted as $f_{G}(g'_{n, K_n})$, can be written as
\begin{equation}
	f_{G}(g'_{n, K_n}) = \frac{\lambda {e^{ - \lambda g'_{n, K_n}}}}{{q(r)}} \cdot \bm{1}(g'_{n, K_n} \geq G(r)).
\end{equation}

For the data offloading in Step 3, a data amount of $D$ nats is required to be offloaded from the winning IoT device to the BS.
For a round, say Round $n$, define the beginning time and ending time of observation offloading as $t_{n,0}$ and $t_{n,1}$ respectively,
then the time span for observation offloading, denoted as $T_{n}$, can be written as
\begin{equation}
	T_n=t_{n,1}-t_{n,0}=\frac{D}{R_{n,K_n}}.
\end{equation}

With the above descriptions, we can see the time span of one round, say Round $n$, is composed of observation offloading time $T_n$ and the competition time of channel accessing right, which is denoted as $P_n$.
Specifically, suppose $J_{n,k}$ time slots are required to produce a winner IoT device for $k$th competition in Round $n$, the $P_n$ can be written as
\begin{equation}
	P_n=\delta \sum \limits_{k=1}^{K_n} J_{n,k}
\end{equation}
Since the contention for the channel accessing are independent Bernoulli processes, the $J_{n,k}$s over disjoint $k$ values are i.i.d random variables following the geometric distribution with parameter being $p_n$, whose probability mass function can be written as $\Pr \left(J_{n,k}=j\right)=(1-p_n(p))^{j-1}p_n(p)$. For brevity, it is  denoted as $J_{n,k} \sim{{Geo}(p_n(p))}$. Similarly, the $K_n$ is also a geometric distributed random variable, but with parameter being $q(r)$, which can be denoted as $K_n \sim{Geo(q(r))}$. 
%which follows the geometric distribution with parameter $q(r)$, denoted as $K_n \sim{GE(q(r))}$.
With the above definitions, 
the time span of one observation span $C$ can be written as $C = \sum_{n=1}^{N} \left(T_n + P_n\right)$.

%\begin{figure}
%\end{figure}

\begin{figure}[!hbtp]
	\begin{center}
		\includegraphics[angle=0,width=0.8 \textwidth]{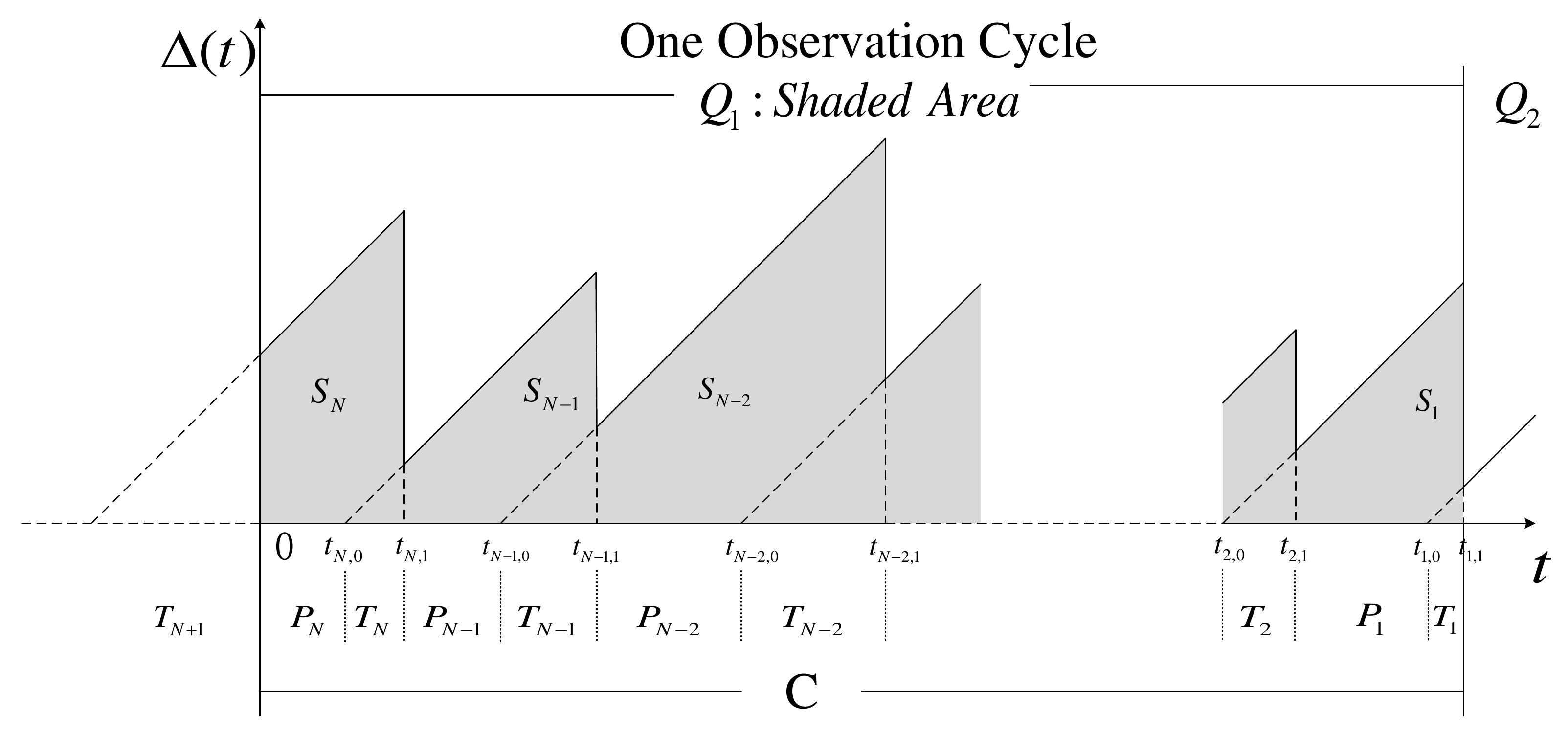}
	\end{center}
	\caption{Example of AoI.}
	\label{fig:AoI}
\end{figure}

Next we are going to investigate the AoI for the described system.
According to \cite{Trends}, the AoI at time instant $t$ is defined as
\begin{equation}
	\Delta(t)=t - u(t)
\end{equation}
where $u(t)$ is the instant of generating the most updated observation at the side of IoT device for a time $t$. The $\Delta(t)$ actually represents the elapsed time since the generation of an observation %before it is fully received by the BS.
%where $u(t)$ is the generation time of the latest update at the base station. It can be described as
%\begin{equation}
%	u(t)=\max{\{t_{n,0}|t_{n,1} \leq t\}}
%\end{equation}
%In an observation cycle, the AoI can be denoted in Fig.\ref{fig:AoI}.
%The whole time of an observation cycle denotes as $C$, where $C=\sum_{n=1}^{N} T_n + P_n$.
Looking into Fig. \ref{fig:AoI}, which plots the instant AoI $\Delta(t)$ as $t$ grows from 0 to $C$. Define $S_n$ as the area of $\Delta(t)$ within time interval of Round $n$ for $n\in \mathcal{N}$, which can be calculated as
\begin{equation}
	{S_n} = \frac{1}{2}\left( {T_{n + 1}} + ({T_{n + 1}} + {P_n} + {T_n}) \right)\left({P_n} + {T_n}\right)
\end{equation}
according to Fig. \ref{fig:AoI} \footnote{When $n=N$, $T_{n+1}$ actually represents the $T_1$ in the last observation cycle.},
then $Q = \sum_{n=1}^{N} S_n$ represents the accumulated AoI in current observation cycle.

To evaluate the average AoI over long time duration that may span multiple observation cycles, we define $Q_m$ as the $Q$ value (i.e., the accumulated AoI) in $m$th observation cycle and $M(T)$ as the number of observation cycles within time $T$, then the averaged AoI over time $T$ can be written as
\begin{equation}
	\overline {\Delta_T}  = \frac{1}{T}\int_0^T {\Delta (t)dt}  = \frac{1}{T}\sum\limits_{m = 1}^{M(T)} {Q_m}  = \frac{M(T)}{T}\frac{1}{M(T)}\sum\limits_{m = 1}^{M(T)} {{Q_m}}
\end{equation}
and the average AoI defined in \cite{Trends} can be written as
\begin{equation}
	\overline \Delta   = \mathop {\lim }\limits_{T \to \infty } \overline {{\Delta _T}}  = \frac{\mathbb{E}\{Q\}}{\mathbb{E}\{C\}}
\end{equation}
considering that $\lim_{T \to \infty} \frac{M(T)}{T} = \frac{1}{\mathbb{E}\{C\}}$.

In this paper, our target is to minimize the average AoI $\overline{\Delta}$. To achieve this goal, both contending probability $p$ and transmit rate threshold $r$ can be optimized, and the following optimization problem is formulated
\begin{prob} \label{prob:all}
\begin{subequations}
\begin{align}
\min \limits_{p,r}~ & \overline \Delta  \nonumber \\
\text{s.t.} \quad
&  0 < p < 1,\label{e:prob_p}\\
&  0 <r \label{e:prob_r}.
\end{align}
\end{subequations}
\end{prob}

\section{Optimal Solution}
To solve Problem \ref{prob:all}, we first look into the objective function $\overline{\Delta}$, which is a joint function of $p$ and $r$, and can be expanded as
\begin{equation} \label{Delta}
	\begin{array}{c}
		\overline \Delta({p,r})  = \frac{\mathbb{E} \left\{\sum\limits_{n = 1}^N {{S_n}} \right\}}{\mathbb{E}\left\{ \sum\limits_{n = 1}^N ({T_n} + {P_n}) \right\} } = \frac{{\sum\limits_{n = 1}^N {\mathbb{E}\left\{{S_n}\right\}} }}{{\sum\limits_{n = 1}^N \big( {\mathbb{E}\left\{{T_n}\right\} +\mathbb{E} \left\{{P_n}\right\}} \big) }}\\
		= \frac{{\sum\limits_{n = 1}^N {\big(\mathbb{E}\left\{{T_{n + 1}}{P_n}\right\} + \mathbb{E}\left\{{T_{n + 1}}{T_n}\right\} + \mathbb{E}\left\{{T_n}{P_n}\right\} + \frac{1}{2}\mathbb{E}\left\{(P_n)^2\right\} + \frac{1}{2}\mathbb{E}\left\{(T_n)^2\right\}\big)} }}{{\sum\limits_{n = 1}^N {\big(\mathbb{E}\left\{{T_n}\right\} + \mathbb{E}\left\{{P_n}\right\}}\big) }} \\
		= \frac{{\sum\limits_{n = 1}^N {\big(\mathbb{E} \left\{P_n\right\} \left(\mathbb{E} \left\{T_n\right\} + \mathbb{E} \left\{T_{n+1}\right\} \right) + \mathbb{E}\left\{{T_{n + 1}}\right\} \mathbb{E}\left\{{T_n}\right\} + \frac{1}{2}\mathbb{E}\left\{(P_n)^2\right\} + \frac{1}{2}\mathbb{E}\left\{(T_n)^2\right\}\big)} }}{{\sum\limits_{n = 1}^N {\big(\mathbb{E}\left\{{T_n}\right\} + \mathbb{E}\left\{{P_n}\right\}}\big) }}
	\end{array}
\end{equation}
since $T_n$, $T_{n+1}$, $P_n$ are independent from each other.

The items in the expression of $\overline{\Delta}(p,r)$ given in (\ref{Delta}) can be derived as follows
\begin{equation}
	\begin{array}{l} \label{tn}
		\mathbb{E}\left\{ {{T_n}} \right\} = \int_G^\infty  {\frac{D}{{B\ln \left(1 + \frac{{{p_T}g{'_{n,{K_n}}}}}{{B{\sigma ^2}}}\right)}} \cdot {f_G}(g{'_{n,{K_n}}})d} g{'_{n,{K_n}}} \\
	\end{array}
\end{equation}
\begin{equation} \label{tn2}
	\begin{array}{l}
	\mathbb{E}\left\{\left(T_n\right)^2\right\} = \int_G^\infty  {\frac{{{D^2}}}{{{{\left( {B\ln \left(1 + \frac{{{p_T}g{'_{n,{K_n}}}}}{{B{\sigma ^2}}}\right)} \right)}^2}}} \cdot {f_G}(g{'_{n,{K_n}}})d} g{'_{n,{K_n}}} \\
    \end{array}
\end{equation}
\begin{equation} {\label{e:Exp_P_n}}
	\begin{comment}
	\begin{array}{l}
		\mathbb{E}\left\{{P_n}\right\} = \mathbb{E}\left\{ {\delta \sum\limits_{k = 1}^{{K_n}} {{J_{n,k}}} } \right\}\\
		=\delta \sum\limits_{m=1}^{\infty} {\left( {P\{ {K_n}   = k_m\} \mathbb{E}\left\{ {\sum\limits_{k = 1}^{k_m} {{J_{n,k}}} } \right\}} \right)} \\
		=\delta \sum\limits_{m=1}^{\infty} {\left( {P\{ {K_n}   = k_m\} \left( {\sum\limits_{k = 1}^{k_m} {\mathbb{E}\left\{{J_{n,k}}\right\}} } \right)} \right)} \\
		%=\delta \sum\limits_{m=1}^{\infty} {\left( {P\{ {K_n} = k_m\}  \times k_m \times \mathbb{E}\left\{{J_{n.k}}\right\}} \right)} \\
		= \delta \mathbb{E}\left\{{J_{n,k}}\right\} \sum\limits_{m=1}^{\infty} {\left( {P\{ {K_n} = k_m\}  \times k_m} \right)} \\
		= \delta \mathbb{E}\left\{{J_{n,k}}\right\}\mathbb{E}\left\{{K_n}\right\} \\
		=\frac{1}{{{p_n}}}\frac{1}{{q(r)}}\delta
	\end{array}
	\end{comment}
%\begin{comment}
     \begin{array}{*{2}{ll}}
     	&\mathbb{E}\left\{{P_n}\right\} &= \mathbb{E}\left\{ {\delta \sum\limits_{k = 1}^{{K_n}} {{J_{n,k}}} } \right\}\\
     	&&=\delta \sum\limits_{m=1}^{\infty} {\left( {P\{ {K_n}   = m\} \mathbb{E}\left\{ {\sum\limits_{k = 1}^{m} {{J_{n,k}}} } \right\}} \right)} \\
     	&&=\delta \sum\limits_{m=1}^{\infty} {\left( {P\{ {K_n}   = m\} \left( {\sum\limits_{k = 1}^{m} {\mathbb{E}\left\{{J_{n,k}}\right\}} } \right)} \right)} \\
     	&&= \delta \mathbb{E}\left\{{J_{n,k}}\right\} \sum\limits_{m=1}^{\infty} {\left( {P\{ {K_n} = m\}  \times m} \right)} \\
     	&&= \delta \mathbb{E}\left\{{J_{n,k}}\right\}\mathbb{E}\left\{{K_n}\right\} \\
     	&&=\frac{1}{{{p_n}(p)}}\frac{1}{{q(r)}}\delta
     \end{array}
%\end{comment}
\end{equation}
\begin{equation} \label{e:Exp_P_n_2}
	\begin{comment}
	\begin{array}{l}
		\mathbb{E}\left\{P_{\rm{n}}^2\right\} = \mathbb{E}\left\{ {{{\left( {\delta \sum\limits_{k = 1}^{{K_n}} {{J_{n,k}}} } \right)}^2}} \right\}\\
		={\delta ^2}\sum\limits_{m=1}^{\infty} {\left( {P\{ {K_n} = k_m\} \mathbb{E}\left\{ {{{\left( {\sum\limits_{k = 1}^{k_m} {{J_{n,k}}} } \right)}^2}} \right\}} \right)} \\
		={\delta ^2}\sum\limits_{m=1}^{\infty} {\left( {P\{ {K_n} = k_m\} \mathbb{E}\left\{ {\sum\limits_{k = 1}^{k_m} {\sum\limits_{k' = 1}^{k_m} {{J_{n,k}}} {J_{n,k'}}} } \right\}} \right)} \\
		= {\delta ^2}\sum\limits_{m=1}^{\infty} {\left( {P\{ {K_n} = k_m\}\left( k_m \mathbb{E}\left\{J_{n,k}^2\right\} +  k_m(k_m - 1){{\left( {\mathbb{E}\left\{ {{J_{n,k}}} \right\}} \right)}^2}\right)} \right)} \\
		= {\delta ^2}\mathbb{E}\left\{J_{n,k}^2\right\}\mathbb{E}\left\{{K_n}\right\} + {\delta ^2}{\left( {\mathbb{E}\left\{ {{J_{n,k}}} \right\}} \right)^2}\mathbb{E}\left\{K_n^2\right\} \\
		\quad- {\delta ^2}{\left( {\mathbb{E}\left\{ {{J_{n,k}}} \right\}} \right)^2}\mathbb{E}\left\{{K_n}\right\} \\
		=\left( {\frac{{2 - q(r)}}{{p_n^2q{{(r)}^2}}} + \frac{{1 - {p_n}}}{{q(r)p_n^2}}} \right){\delta ^2}
	\end{array}
	\end{comment}
%	\begin{comment}
	\begin{array}{*{2}{ll}}
	&\mathbb{E}\left\{\left(P_n\right)^2\right\} &= \mathbb{E}\left\{ {{{\left( {\delta \sum\limits_{k = 1}^{{K_n}} {{J_{n,k}}} } \right)}^2}} \right\}\\
	&&={\delta ^2}\sum\limits_{m=1}^{\infty} {\left( {P\{ {K_n} = m\} \mathbb{E}\left\{ {{{\left( {\sum\limits_{k = 1}^{m} {{J_{n,k}}} } \right)}^2}} \right\}} \right)} \\
	&&={\delta ^2}\sum\limits_{m=1}^{\infty} {\left( {P\{ {K_n} = m\} \mathbb{E}\left\{ {\sum\limits_{k = 1}^{m} {\sum\limits_{k' = 1}^{m} {{J_{n,k}}} {J_{n,k'}}} } \right\}} \right)} \\
	&&= {\delta ^2}\sum\limits_{m=1}^{\infty} {\left( {P\{ {K_n} = m\}\left( m \mathbb{E}\left\{\left(J_{n,k}\right)^2\right\} +  m(m - 1){\mathbb{E}^2\left\{ {{J_{n,k}}} \right\}}\right)} \right)} \\
	&&= {\delta ^2}\mathbb{E}\left\{\left(J_{n,k}\right)^2\right\}\mathbb{E}\left\{{K_n}\right\} + {\delta ^2}{ {\mathbb{E}^2\left\{ {{J_{n,k}}} \right\}} }\mathbb{E}\left\{\left(K_n\right)^2\right\} 
	- {\delta ^2}{\mathbb{E}^2\left\{ {{J_{n,k}}} \right\}}\mathbb{E}\left\{{K_n}\right\} \\
	&&=\left( {\frac{{2 - q(r)}}{{{p_n(p)}^2q{{(r)}^2}}} + \frac{{1 - {p_n}(p)}}{{q(r){p_n(p)}^2}}} \right){\delta ^2}
	\end{array}
%	\end{comment}
\end{equation}
where the last equality in (\ref{e:Exp_P_n}) and (\ref{e:Exp_P_n_2}) hold since $\mathbb{E}\left\{{K_n}\right\} = \frac{1}{{q(r)}}$,  $\mathbb{E}\left\{\left(K_n\right)^2\right\} = \frac{{2 - q(r)}}{{q{{(r)}^2}}}$, $\mathbb{E}\left\{{J_{n,k}}\right\} = \frac{1}{{{p_n}(p)}}$, and $\mathbb{E}\left\{\left(J_{n,k}\right)^2\right\} = \frac{{2 - {p_n(p)}}}{{{p_n(p)}^2}}$, considering that $J_{n,k} \sim{Geo(p_n(p))}$ and $K_n \sim{Geo(q(r))}$.

Since the objective function $\overline{\Delta}(p,r)$ derived in (\ref{Delta}) is in fractional form, Dinkelbach algorithm can be leveraged to minimize it \cite{dinkelbach}. 
{
Specifically, define function $F(\xi,p,r)$ as
\begin{small}
\begin{equation} \label{sub form}
	\begin{array}{*{2}{ll}}
	&F(\xi,p,r )  \buildrel \Delta \over =&  
	  \sum\limits_{n = 1}^N \Big(\mathbb{E}\left\{{T_{n + 1}}\right\}\mathbb{E}\left\{{P_n}\right\} 
	  + \mathbb{E}\left\{{T_{n + 1}}\right\}\mathbb{E}\left\{{T_n}\right\} 
	  + \mathbb{E}\left\{{T_n}\right\}\mathbb{E}\left\{{P_n}\right\}  
	+ \frac{1}{2}\mathbb{E}\left\{\left(P_n\right)^2\right\} + \frac{1}{2}\mathbb{E}\left\{\left(T_n\right)^2\right\}\Big)  \\&&-\xi\sum\limits_{n = 1}^N \Big(   {\mathbb{E}\left\{{T_n}\right\} + \mathbb{E}\left\{{P_n}\right\}}  \Big) \\
	&&{=}\sum\limits_{n = 1}^N \Big(2\mathbb{E}\left\{{T_{n}}\right\}\mathbb{E}\left\{{P_n}\right\} + \mathbb{E}^2\left\{{T_n}\right\}   
	+ \frac{1}{2}\mathbb{E}\left\{\left(P_n\right)^2\right\} + \frac{1}{2}\mathbb{E}\left\{\left(T_n\right)^2\right\}\Big)  -\xi\sum\limits_{n = 1}^N \Big(   {\mathbb{E}\left\{{T_n}\right\} + \mathbb{E}\left\{{P_n}\right\}}  \Big), 
    \end{array}
\end{equation}
\end{small} 
where the second equality of (\ref{sub form}) holds since $\mathbb{E}\left\{{T_n}\right\}=\mathbb{E}\left\{{T_{n + 1}}\right\}$ according to  (\ref{tn}), and define $F(\xi)$ as 
}
\begin{prob} \label{prob:upper}
	\begin{equation}
		\begin{array}{ll}
			  \Gamma(\xi)  \triangleq & \min \limits_{p,r}~ F(\xi,p,r )   \\
			\text{s.t.} 
			&(\ref{e:prob_p}), (\ref{e:prob_r})
		\end{array}
	\end{equation}
\end{prob}
where the optimal solution of $p$ and $r$ for Problem \ref{prob:upper} are denoted as $p^*(\xi)$ and $r^*(\xi)$ respectively. 
%where $p$ and $r$ in (\ref{sub form}) (\ref{sub form2}) are subject to (\ref{e:prob_p})(\ref{e:prob_r}). 

According to \cite{dinkelbach}, the minimal achievable cost function of Problem \ref{prob:all} would be the $\xi^*$ such that $\Gamma(\xi^*)=0$, and the optimal solution of $p$ and $r$ for Problem \ref{prob:all} would be exactly the optimal solution for Problem \ref{prob:upper} when $\xi = \xi^*$, i.e., $p^*(\xi^*)$ and $r^*(\xi^*)$. 
Then an iterative search method can be employed to find the $\xi^*$. 
In $i$th iteration, the current $\xi$ is denoted as $\xi_i$, we need to first work out the $p^*(\xi_i)$ and $r^*(\xi_i)$ respectively by solving Problem \ref{prob:upper}, and then calculate the $\xi$ for $(i+1)$th iteration, which is given as $\xi_{i+1}=\overline \Delta\left({p^*(\xi_i), r^*(\xi_i)}\right)$.
This iteration will stop until the most updated $\xi$ approaches to $\xi^*$.

%For such an optimal Problem, the optimal solution $(p^\star(i),r^\star(q))$ can be calculated in terms of the following Lemmas.
For Problem \ref{prob:upper}, it can be checked the objective function is non-convex with $(p, r)^T$, which brings challenge into problem solving. To find the solution of Problem \ref{prob:upper}, we utilize the Block Coordinate Descent (BCD) method to optimize $p$ and $r$ iteratively.
Specifically, starting from a feasible solution of Problem \ref{prob:upper} for $\xi = \xi_i$, denoted as $\{p_0(\xi_i),r_0(\xi_i)\}$, $r_{j}(\xi_i)$, we optimize $p$ to obtain $p_j(\xi)$ with $r$ fixed at $r_{j-1}(\xi_i)$ for Problem \ref{prob:upper} and then optimize $r$ to obtain $r_j(\xi)$ with $p$ fixed at $p_{j}(\xi_i)$ for Problem \ref{prob:upper}. This process is performed iteratively until the convergence of Problem  \ref{prob:upper}'s objective function.

In the BCD solving procedure, although the optimization of $p$ or $r$ only involves the minimization of a single-variable function (i.e., the objective function of Problem \ref{prob:upper} with $r$ or $p$), recalling the complicated expression of $\mathbb{E}\left\{{T_n}\right\}$, $\mathbb{E}\left\{\left(T_n\right)^2\right\}$, and $\mathbb{E}\left\{\left(P_n\right)^2\right\}$, it is hard to find the optimal solution of $p$ or $r$ by simply setting the derivative of the Problem \ref{prob:upper}'s objective function to be zero. To work out the optimal $p$ or $r$ in each iterative step of BCD method, we have to explore some other way.

In term of Problem \ref{prob:upper}'s objective function with respect to $r$, the following lemmas can be anticipated
%Hence we analyze the monotonicity of objective function with respect to $r$ and have the following property. 
\begin{lemma} \label{lem:mono_r}
In Problem \ref{prob:upper}'s objective function
	%The monotonicity of each part of the objective function with respect to $r$ is as follows,
	\begin{itemize}
		\item The terms $\frac{1}{2}\mathbb{E}\left\{\left(P_n\right)^2\right\}$, $-\xi \mathbb{E}\left\{{T_n}\right\}$ are all monotonically increasing with $r$;
		{
		\item The terms $\mathbb{E}^2\left\{{T_n}\right\}$, $\frac{1}{2}\mathbb{E}\left\{\left(T_n\right)^2\right\}$, $- \xi \mathbb{E}\left\{{P_n}\right\}$ are monotonically decreasing with $r$.
	    }
	\end{itemize}
\end{lemma}
\begin{IEEEproof}
Please refer to Appendix \ref{proof monotonical r}.
\end{IEEEproof}
{	
	\begin{lemma} \label{lem:mono_r etnpn}
	$\mathbb{E}\left\{{T_n}\right\}\mathbb{E}\left\{{P_n}\right\}$ is a unimodal function with $r$. Specifically, there exists a point $r^\ddagger$ such that $\mathbb{E}\left\{{T_n}\right\}\mathbb{E}\left\{{P_n}\right\}$ is monotonically decreasing for $0< r \leq r^\ddagger$ and monotonically increasing for $r> r^\ddagger$. 
\end{lemma}
\begin{IEEEproof}
Please refer to Appendix \ref{proof_B}.
\end{IEEEproof}
}

From the proof of Lemma \ref{lem:mono_r etnpn}, it is hard to characterize the $r^{\ddagger}$. On the other hand, 
we can find the $r^{\ddagger}$ through performing bisection search of $r$ to achieve the minimal value of $\mathbb{E}\left\{{T_n}\right\}\mathbb{E}\left\{{P_n}\right\}$, since the term $\mathbb{E}\left\{{T_n}\right\}\mathbb{E}\left\{{P_n}\right\}$ is a unimodal function that first decreases and then increases, which is disclosed by Lemma \ref{lem:mono_r etnpn}.
Collecting the results of Lemma \ref{lem:mono_r} and Lemma \ref{lem:mono_r etnpn}, 
Problem \ref{prob:upper}'s objective function can be written as the difference of two monotonically increasing functions with $r$ for $r \in (0, r^{\ddagger}]$, and can be written as difference of another two monotonically increasing functions for $r \in (r^{\ddagger}, \infty)$, the expressions of which can be found in (\ref{r divide}).
%\begin{figure*}
\begin{small}
\begin{equation} \label{r divide}
	\begin{split}
&F(\xi,p,r ) = \\ &\left\{
   \begin{array}{*{2}{ll}}
\sum\limits_{n = 1}^N \left(\frac{1}{2}\mathbb{E}\left\{\left(P_n\right)^2\right\} -\xi \mathbb{E}\left\{{T_n}\right\} \right)  
- \sum \limits_{n=1}^N - \Big( 2 \mathbb{E}\left\{{T_{n}}\right\} \mathbb{E}\left\{{P_n}\right\} +\mathbb{E}^2\{T_n\}  
+ \frac{1}{2}\mathbb{E}\left\{\left(T_n\right)^2\right\}  - \xi \mathbb{E}\left\{{P_n}\right\} \Big),  &0<r\leq r^\dagger \\
\sum\limits_{n = 1}^N \left(\frac{1}{2}\mathbb{E}\left\{\left(P_n\right)^2\right\} -\xi \mathbb{E}\left\{{T_n}\right\} +2\mathbb{E}\left\{{T_{n}}\right\} \mathbb{E}\left\{{P_n}\right\} \right)  - \sum \limits_{n=1}^N - \Big(  \mathbb{E}^2\{T_n\}
+ \frac{1}{2}\mathbb{E}\left\{\left(T_n\right)^2\right\}  - \xi \mathbb{E}\left\{{P_n}\right\} \Big), & r>r^\dagger 
	\end{array}
 \right.
\end{split}
\end{equation}
\end{small}
%\hrulefill
%\end{figure*}

As of minimizing the difference of two monotonically increasing functions for $r\in (0, r^{\ddagger}]$ or $r \in (r^{\ddagger}, \infty)$, it can be transformed to be standard Monotonic optimization problem equivalently \cite{monotonic}, whose global optimal solution is able to be found through Polyblock algorithm \cite{monotonic}.
Due to the limit of space, the detail is omitted here.
In the final step, the global optimal $r$ to achieve minimal value of Problem \ref{prob:upper}'s objective function can be obtained by selecting the better one for minimizing Problem \ref{prob:upper}'s objective function for $r\in (0, r^{\ddagger}]$ and $r \in (r^{\ddagger}, \infty)$.

In terms of  Problem \ref{prob:upper}'s objective function with respect to $p$, for the ease of discussion, we collect the items related to $n$th IoT device in the expression of $F(\xi, p, r)$ as $F_n(\xi, p, r)$, then there is $F(\xi,p,r)=\sum_{n = 1}^NF_n(\xi,p,r)$ and $F_n(\xi, p, r)$ can be expressed as 
\begin{equation}
	 F_n(\xi,p,r) = A_n \left({\frac{1}{{{p_n(p)}}}}\right)^2 + \left(B_n + C_n \right) \frac{1}{{{p_n(p)}}} + D_n
\end{equation}
where 
${A_n} = \frac{{{\delta ^2}}}{{q{{(r)}^2}}}$, ${B_n} = \frac{{2\delta\mathbb{E}\left\{ {T_n} \right\} }}{{q(r)}}$, ${C_n} =  - \frac{{{\delta ^2}}}{{2q(r)}} - \xi \frac{\delta }{{q(r)}}$, ${D_n} = \frac{1}{2}\mathbb{E}\left\{ {{\left(T_n\right)}^2} \right\} + \mathbb{E}^2\{T_n\} - \xi \mathbb{E}\left\{ {T_n} \right\}$. 
We further decompose $F_n(\xi, p, r)$ as $F_n(\xi,p,r) = F_{n,1}(\xi,p,r) + F_{n,2}(\xi,p,r)$, such that 
$F_{n,1}(\xi,p,r) \triangleq A_n \cdot \left({\frac{1}{{{p_n(p)}}}}\right)^2 + B_n \cdot \frac{1}{{{p_n(p)}}}$ and
$F_{n,2}(\xi,p,r) \triangleq C_n \cdot \frac{1}{{{p_n(p)}}} +D_n$.

It can be checked that $F_{n,1}(\xi,p,r)$ is decreasing with $p_n(p)$ and $F_{n,2}(\xi,p,r)$ is increasing with $p_n(p)$.
Since $p_n(p)=n p (1-p)^{n-1}$ increases monotonically in for $p \in [0,1/n]$ and decreases for $p \in [1/n,1]$, we can summarize that
%\begin{lemma} \label{unimodal}
\begin{itemize}
	\item For $p \in (0,1/n)$, $F_{n,1}(\xi,p,r)$ is monotonically decreasing, and $F_{n,2}(\xi,p,r)$ is monotonically increasing with respect to $p$.
	\item For $p \in [1/n,1)$, $F_{n,1}(\xi,p,r)$ is monotonically increasing, and $F_{n,2}(\xi,p,r)$ is monotonically decreasing with respect to $p$.
\end{itemize}	
%\end{lemma}
Then we can divide the feasible interval of $p$ $(0, 1)$ into $N$ disjoint sub-intervals $(0, 1/N)$, $[1/N, 1/(N-1))$, ..., $[1/2, 1)$. For these sub-intervals, there is 
%Thus, for the objective function $F(\xi,p,r)$, there are
%\begin{lemma} \label{interval}
	\begin{itemize}
		\item For $p \in (0,1/N)$, $\sum_{n = 1}^N F_{n,1}(\xi,p,r)$ is decreasing and  $\sum_{n = 1}^N F_{n,2}(\xi,p,r)$ is increasing with respect to $p$.
		\item For $p \in [\frac{1}{n''},\frac{1}{n''-1})$,
		where $n'' \in \{2, ..., N\}$, $\sum_{n = 1}^{n''-1} F_{n,1}(\xi,p,r)+\sum_{n = n''}^{N} F_{n,2}(\xi,p,r)$ is decreasing and $\sum_{n = 1}^{n''-1} F_{n,2}(\xi,p,r)+\sum_{n = n''}^{N} F_{n,1}(\xi,p,r)$ is increasing with respect to $p$.
	\end{itemize}
%\end{lemma}
Recalling the fact that $F(\xi,p,r)=\sum_{n = 1}^NF_n(\xi,p,r) = \sum_{n=1}^N \left(F_{n,1}(\xi,p,r) + F_{n,2}(\xi,p,r)\right)$, we can claim that in each sub-interval of $p$, $F(\xi, p, r)$ can be always written as the difference of two increasing functions of $p$, the minimal value of $F(\xi,p,r)$ in this interval can be searched by the polyblock algorithm as we mention for the optimization of $r$.
Then the global minimal value of $F(\xi, p, r)$ can be found by selecting the sub-interval associated with the least minimal value of $F(\xi, p, r)$.
To this end, how to work out the optimal $p$ so as to minimize $F(\xi, p, r)$ has been completed.

\section{Numerical Results} \label{s:num}
\begin{figure}[!hbtp]
	\begin{center}
		\includegraphics[angle=0,width=0.55 \textwidth]{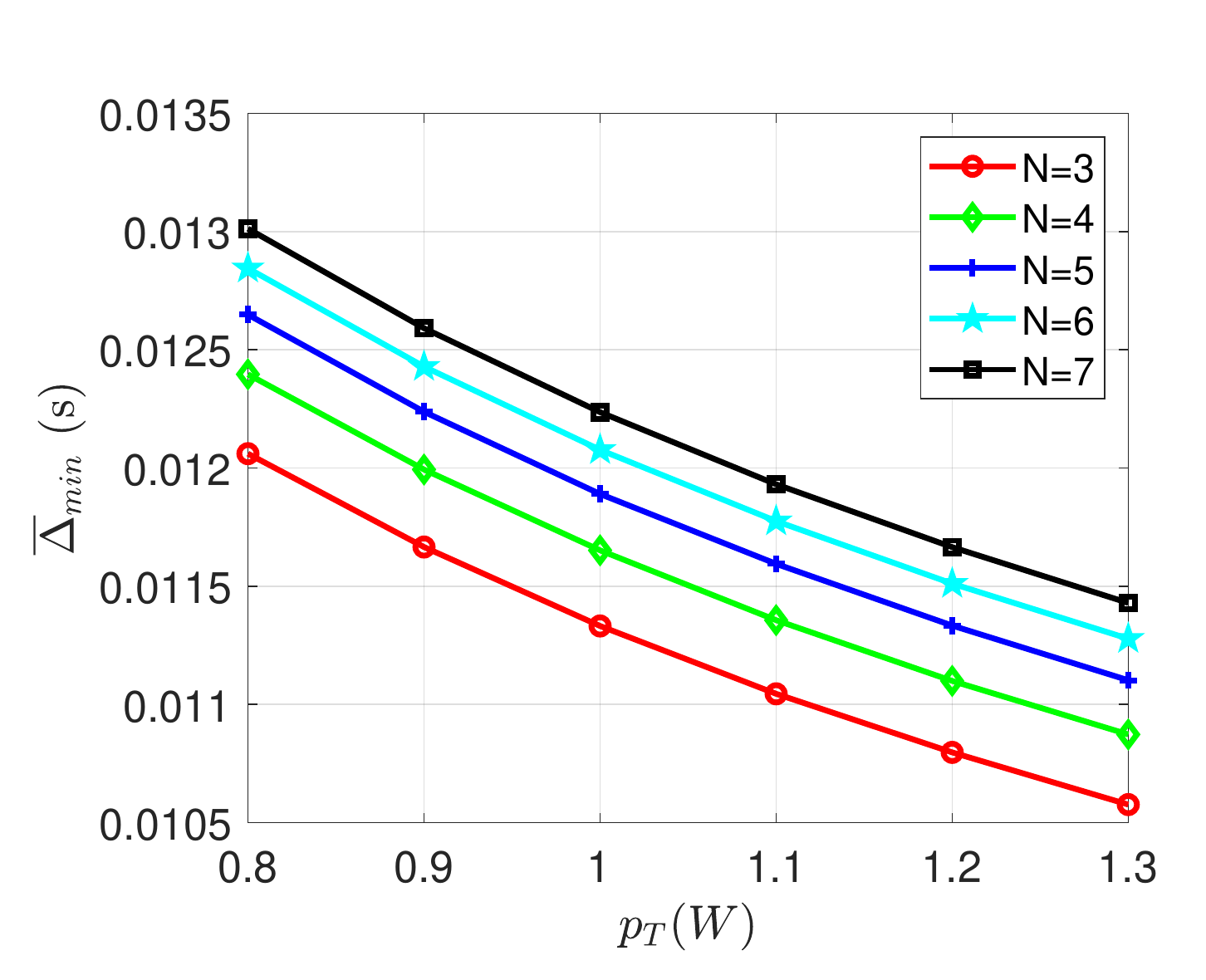}
	\end{center}
	\caption{Minimum average AoI versus $p_T$ for $N=\{3,4,5,6,7\}$}
	\label{fig:AoI_result}
\end{figure}

\begin{figure}[!hbtp]
	\begin{center}
		\includegraphics[angle=0,width=0.5 \textwidth]{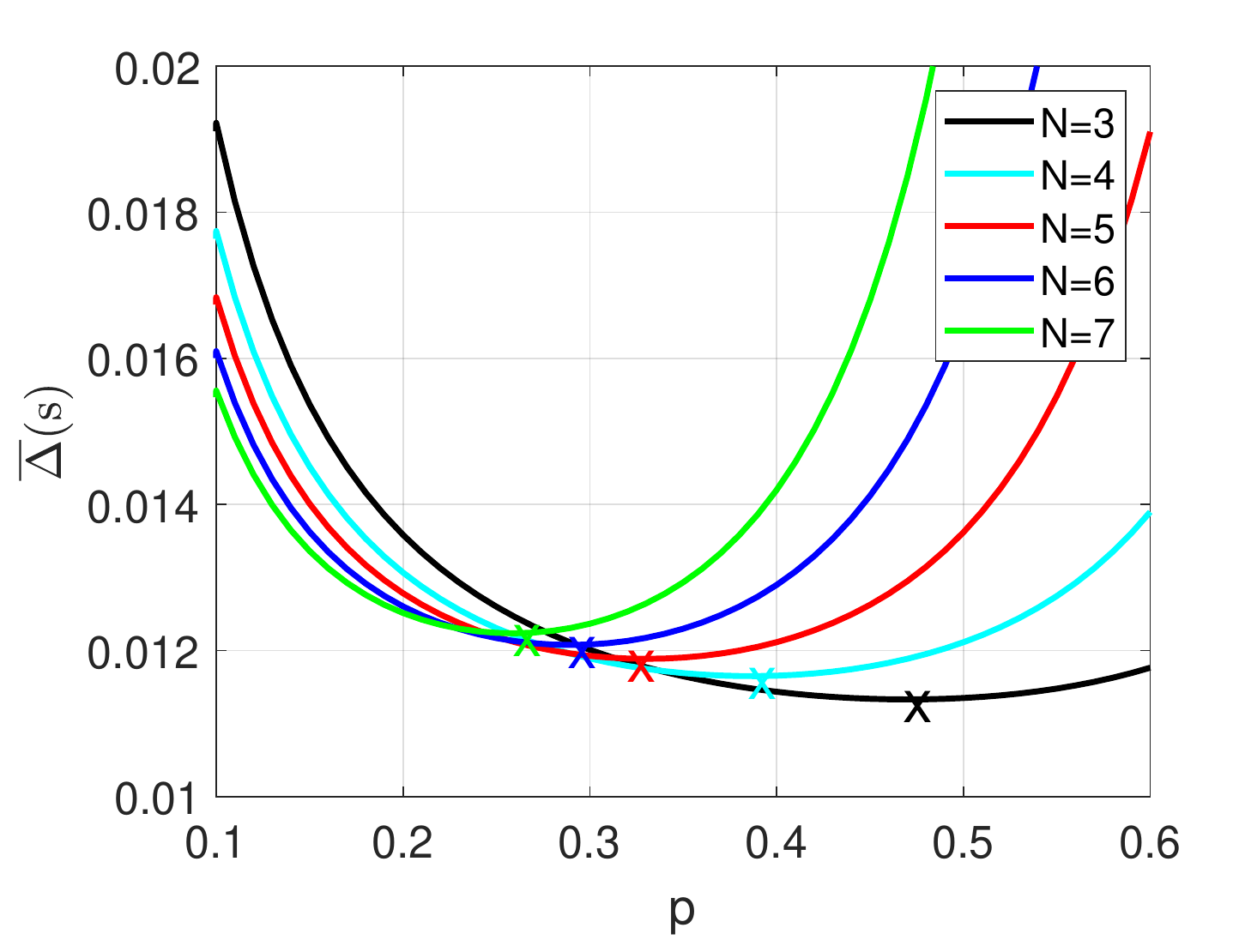}
	\end{center}
	\caption{$\Delta$ and Optimal $p^*$ with $p_T=1W$ and $N=\{3,4,5,6,7\}$}
	\label{fig:vs_p}
\end{figure}

In this section, numerical results are presented to analyze the performance of our proposed method.
The system parameters are set as follows in default.
The distance between devices and the BS is $500$ m. The channel noise spectrum density is $\sigma^2=-140$dBm, the bandwidth is $B=1$MHz, the carrier frequency is $4.8$ GHz and the mean of random Rayleigh fading is $1$.
The free space path loss, which is denoted as $l_f$, can be calculated from the following formula (in dB)
\begin{equation}
	l_f=32.4+20\times\log(Distanse)+20\times\log(Frequency)
\end{equation}
The length of one time slot $\delta=0.001$s and the amount of data for every update $D=1.5\times10^4$ nats \cite{data}.

Fig. \ref{fig:AoI_result} plots the minimum average AoI versus $p_T$ for different number of devices. As it is expected, for a given $p_T$, a bigger $N$ increases the achieved AoI, as there are more collisions before a successful transmission take place.

Fig. \ref{fig:vs_p} depicts the average AoI versus the contention probability $p$ with the optimal $r^*$ for different number of devices and a certain $p_T$. Specifically, a large $p$ causes more collisions in the front of an observation cycle where there are more devices. But a small $p$ causes more idle at the end of the cycle where there are few devices. So the optimal $p^*$ is different for each $N$. And for each $N$, the optimal $p^*$ which is calculated by our proposed solution matches the lowest point of each curve and verifies the validity of our proposed method.

%Fig. \ref{fig:comparison} depicts the minimum average AoI results from different methods. We compare the method which is proposed in \cite{FDMA}, i.e. golden-search method, with monotonic optimization algorithm in the optimization for $r$ and $p$. We set the same convergence precision in two different algorithm, and keep the settings of Dinkelbach algorithm and BCD algotithm the same. It is obvious that our proposed method has better performance.

\section{Conclusion} \label{s:conclusion}
In this paper, we have investigated the performance of an IoT system with opportunistic channel access in terms of the average AoI. 
%Devices transmit updates to the BS by contending to access channel and comparing it with the transmission rate threshold. 
In order to minimize the average AoI of this system, the contention probability and transmission rate threshold are optimized.
Although the original problem with fractional form objective function is complex, by decomposing the associated problem into two single-variable sub-problems through Dinkelbach method and BCD method, the optimal solution can be found by Monotonic optimization method.
%We derived the expressions for the average AoI with some random variables, which depend on contention probability and transmission rate threshold. 
%And we verified the independence and probability distribution of these variables and verified the monotonicity of AoI for each optimization variable. 
%The optimal contention probability and transmission rate threshold have been calculated numerically by formulating and solving a two-dimensional optimization problem.
Numerical results verifies the effectiveness of our proposed method. This research could provide helpful insight on improving the information freshness in multiple devices IoT system with opportunity channel access.

\appendices

\section{The proof of Lemma \ref{lem:mono_r}} \label{proof monotonical r}
We first look into the monotonicity of $\mathbb{E}\left\{T_n\right\}$ with respect to $r$.
The $\mathbb{E}\left\{ {T_n} \right\}$ can be written as
\begin{equation}	
			\begin{aligned}
				\mathbb{E}\left\{ {T_n} \right\} &= \int_{G(r)}^\infty  {\frac{D}{{B\ln \left(1 + \frac{{{p_T}g{'_{n,{K_n}}}}}{{B{\sigma ^2}}}\right)}} \cdot  {f_G}(g{'_{n,{K_n}}})d} g{'_{n,{K_n}}}\\
%				& = \int_{G(r)}^\infty  {\frac{D}{{B\ln \left(1 + \frac{{{P_T}g{'_{n,{K_n}}}}}{{B{\sigma ^2}}}\right)}} \times \frac{{\lambda {e^{ - \lambda {{g'}_{n,{K_n}}}}}}}{{q(r)}}d} g{'_{n,{K_n}}}\\
				& = \int_{G(r)}^\infty  {\frac{D}{{B\ln \left(1 + \frac{{{p_T}g{'_{n,{K_n}}}}}{{B{\sigma ^2}}}\right)}} \times \frac{{\lambda {e^{ - \lambda {{g'}_{n,{K_n}}}}}}}{{\int_{G(r)}^\infty  {\lambda {e^{ - \lambda {g_{n,i}}}}d{g_{n,i}}} }}d} g{'_{n,{K_n}}}\\
				&= \frac{{\int_{G(r)}^\infty  {\frac{{D\lambda {e^{ - \lambda {{g'}_{n,{K_n}}}}}}}{{B\ln \left(1 + \frac{{{p_T}g{'_{n,{K_n}}}}}{{B{\sigma ^2}}}\right)}}d} g{'_{n,{K_n}}}}}{{\int_{G(r)}^\infty  {\lambda {e^{ - \lambda {g_{n,i}}}}d{g_{n,i}}} }}
			\end{aligned}
\end{equation}
The partial derivative of $\mathbb{E} \left\{T_n\right\}$ with $G(r)$ is given as
\begin{equation} \label{e:dT_dG}
	\begin{aligned}
	\frac{{\partial \mathbb{E}\left\{ {{T_n}}\right\}} }{{\partial G(r)}} &=  
	\frac{{\int_{G(r)}^\infty  {\frac{{D\lambda {e^{ - \lambda {{g'}_{n,{K_n}}}}} \times \lambda {e^{ - \lambda {G(r)}}}}}{{B\ln \left(1 + \frac{{{p_T}g{'_{n,{K_n}}}}}{{B{\sigma ^2}}}\right)}}d} g{'_{n,{K_n}}}  d {g_{n,i}}}}{{{{\left[ {\int_{G(r)}^\infty  {\lambda {e^{ - \lambda {g_{n,i}}}}d{g_{n,i}}} } \right]}^2}}} -  
	 \frac{{  - \int_{G(r)}^\infty  {\frac{{D\lambda {e^{ - \lambda {g_{n,i}}}} \times \lambda {e^{ \lambda {G(r)}}}}}{{B\ln \left(1 + \frac{{{p_T}{G(r)}}}{{B{\sigma ^2}}}\right)}}d} {g_{n,i}}}}{{{{\left[ {\int_{G(r)}^\infty  {\lambda {e^{ - \lambda {g_{n,i}}}}d{g_{n,i}}} } \right]}^2}}}\\
		&= \frac{{\int_{G(r)}^\infty  {\frac{{D\lambda {e^{ - \lambda g}} \times \lambda {e^{ - \lambda G(r)}}}}{{B\ln \left(1 + \frac{{{p_T}g}}{{B{\sigma ^2}}}\right)}}d} g - \int_{G(r)}^\infty  {\frac{{D\lambda {e^{ - \lambda g}} \times \lambda {e^{ - \lambda {G(r)}}}}}{{B\ln \left(1 + \frac{{{p_T}{G(r)}}}{{B{\sigma ^2}}}\right)}}d} g}}{{{{\left[ {\int_{G(r)}^\infty  {\lambda {e^{ - \lambda {g_{n,i}}}}d{g_{n,i}}} } \right]}^2}}}\\
		&= \frac{{\int_{G(r)}^\infty  {\left( \frac{{D\lambda {e^{ - \lambda g}} \times \lambda {e^{ - \lambda {G(r)}}}}}{{B\ln \left(1 + \frac{{{p_T}g}}{{B{\sigma ^2}}}\right)}} - \frac{{D\lambda {e^{ - \lambda g}} \times \lambda {e^{ - \lambda {G(r)}}}}}{{B\ln \left(1 + \frac{{{p_T} {G(r)}}}{{B{\sigma ^2}}}\right)}}\right) d} g}}{{{{\left[ {\int_{G(r)}^\infty  {\lambda {e^{ - \lambda {g_{n,i}}}}d{g_{n,i}}} } \right]}^2}}}
   \end{aligned}
\end{equation}
Since the $g \ge {G(r)}$ in (\ref{e:dT_dG}), it can be checked that $\frac{{\partial \mathbb{E}\left\{ {{T_n}}\right\}} }{{\partial {G(r)}}} \leq 0$, which implies the decreasing monotonicity of $T_n$ with ${G(r)}$. 
%Because of $g \ge G$, The partial derivative must be non-positive, and $\mathbb{E}\left\{ {T_n} \right\}$ is decreasing for $G$.
 Recalling that ${G(r)}$ is increasing for $r$, it can be proved that $\mathbb{E}\left\{T_n\right\}$ is decreasing with $r$. 
Similarly, the monotonicity of $\mathbb{E}^2\{T_n\}$, $\mathbb{E}\left\{{\left(T_n\right)}^2\right\}$,  $\mathbb{E}\left\{P_n\right\}$ and $\mathbb{E}\left\{\left(P_n\right)^2\right\}$ with respect to $r$ can be also proved.
%Similarly, the monotonicity of $\mathbb{E}\left\{P_n\right\}$ and $\mathbb{E}\left\{P^2_n\right\}$ can be proved by taking derivatives, which both increasing for $r$. 
At this point, the monotonicity of $\frac{1}{2}\mathbb{E}\left\{\left(P_n\right)^2\right\}$, $- \xi \mathbb{E}\left\{{T_n}\right\}$, $\mathbb{E}^2\left\{{T_n}\right\}$, $\frac{1}{2}\mathbb{E}\left\{\left(T_n\right)^2\right\}$, $- \xi \mathbb{E}\left\{{P_n}\right\}$ with respect to $r$ has been proved.

{
\section{The proof of Lemma \ref{lem:mono_r etnpn}} \label{proof_B}
We first inspect $\mathbb{E}\left\{{T_n}\right\}\mathbb{E}\left\{{P_n}\right\}$ and its partial derivative with $G(r)$, which can be written as 
\begin{equation}
	\begin{array}{*{2}{ll}}
	&\mathbb{E}\left\{{T_n}\right\}\mathbb{E}\left\{{P_n}\right\}&= 
	\int_{G(r)}^\infty  {\frac{D}{{B\ln \left(1 + \frac{{{p_T}g{'_{n,{K_n}}}}}{{B{\sigma ^2}}}\right)}}  {f_G}(g{'_{n,{K_n}}})d} g{'_{n,{K_n}}}  \cdot \frac{{\delta }}{{{p_n(p)}{q(r)}}} \\
	 &&= \frac{\delta }{{{p_n(p)}}} \cdot {{\int_{G(r)}^\infty  {\frac{{D\lambda {e^{ - \lambda {{g'}_{n,{K_n}}}}}}}{{B\ln \left(1 + \frac{{{p_T}g{'_{n,{K_n}}}}}{{B{\sigma ^2}}}\right)}}d} g{'_{n,{K_n}}}}}/
	 {{\int_{G(r)}^\infty  {2\lambda {e^{ - 2\lambda {g_{n,i}}}}d{g_{n,i}}} }},
    \end{array}
\end{equation}
and
\begin{equation}\label{one moment}
	\begin{array}{l}
	\frac{{\partial \left(\mathbb{E}\left\{{T_n}\right\}\mathbb{E}\left\{{P_n}\right\}\right)}}{\partial {G(r)}} 
		=\frac{{2D{\lambda ^3}\delta }}{{B{p_n}(p)}}\frac{{\frac{{{e^{ - \lambda G(r)}}}}{{\ln \left( {1 + \frac{{{p_T}G(r)}}{{B{\sigma ^2}}}} \right)}}\int_{G(r)}^\infty  {\frac{{{e^{ - \lambda g}}}}{{\ln \left( {1 + \frac{{{p_T}g}}{{B{\sigma ^2}}}} \right)}}} \left( {\frac{{\ln \left( {1 + \frac{{{p_T}G(r)}}{{B{\sigma ^2}}}} \right)}}{{{e^{\lambda G(r)}}}} - \frac{{\ln \left( {1 + \frac{{{p_T}g}}{{B{\sigma ^2}}}} \right)}}{{{e^{\lambda g}}}}} \right)dg}}{{{{\left[ {\int_{G(r)}^\infty  {2\lambda {e^{ - 2\lambda {g_{n,i}}}}d{g_{n,i}}} } \right]}^2}}}.
	\end{array}
\end{equation}
In order to check the polarity of $\frac{{\partial \left(\mathbb{E}\left\{{T_n}\right\}\mathbb{E}\left\{{P_n}\right\}\right)}}{\partial {G(r)}}$, which also reflects the monotonicity of $\mathbb{E}\left\{{T_n}\right\}\mathbb{E}\left\{{P_n}\right\}$ with $G(r)$, we define an auxiliary functions ${G_a}(x) = \frac{{\ln \left( {1 + \frac{{{p_T}x}}{{B{\sigma ^2}}}} \right)}}{{{e^{\lambda x}}}}$, then there is 
\begin{equation}\label{monotonic auxiliary}
	\frac{{\partial {G_a}(x)}}{{\partial x}} = \frac{{\frac{{{p_T}}}{{B{\sigma ^2} + {p_T}x}} - \lambda \ln \left( {1 + \frac{{{p_T}x}}{{B{\sigma ^2}}}} \right)}}{{{e^{\lambda x}}}}.
\end{equation}
It is obvious that for $x > 0$, $\frac{{\partial {G_a}(x)}}{{\partial x}}$ has only one zero point, which can be denoted as $x^\dagger$. Hence it can be inferred that $G_a(x)$ increases monotonically for $ x\in (0,x^\dagger]$ and decreases for $ x\in (x^\dagger,+\infty)$.
}

{ 
Back to (\ref{one moment}), when $G(r) > x^\dagger$, for $g \geq G(r)$, there is $G_a(G(r)) \geq G_a(g)$ since $G_a(x)$ is monotonically decreasing for $x > x^{\dagger}$. 
%According to (\ref{one moment}), we can prove that for $G(r) > x^\dagger$,
Then there is 
\begin{equation}\label{partial 0}
 \frac{{\partial \left(\mathbb{E}\left\{{T_n}\right\}\mathbb{E}\left\{{P_n}\right\}\right)}}{\partial {G(r)}} > 0, \forall G(r) > x^{\dagger}.
\end{equation}
}

{
Next we switch to another case, $G(r) \in (0, x^{\dagger}]$.
In this case,  it is hard to check the polarity of (\ref{one moment}) directly. Alternatively, 
we rewrite (\ref{one moment}) in another form
\begin{equation}\label{F1}
	\begin{array}{l}
	\frac{{\partial \left(\mathbb{E}\left\{{T_n}\right\}\mathbb{E}\left\{{P_n}\right\}\right)}}{\partial {G(r)}} 
	 = \frac{{D{\lambda ^2}\delta }}{{B{p_n}(p)}}{e^{2\lambda G(r)}}\left( {2\lambda \int_{G(r)}^\infty  {\frac{{{e^{ - \lambda g}}}}{{\ln \left( {1 + \frac{{{p_T}g}}{{B{\sigma ^2}}}} \right)}}} dg{\rm{ - }}\frac{{{e^{ - \lambda G(r)}}}}{{\ln \left( {1 + \frac{{{p_T}G(r)}}{{B{\sigma ^2}}}} \right)}}} \right)
	 \end{array}
\end{equation}
Since $\frac{{D{\lambda ^2}\delta }}{{B{p_n}(p)}}{e^{2\lambda G(r)}}$ in (\ref{F1}) is always positive for $G(r)>0$, 
we only need to focus on the rest part of the right-hand side of (\ref{F1}) in terms of its polarity, which can be denoted as
\begin{equation} \label{F11}
	\Xi_{(1)}(G(r))= 2\lambda \int_{G(r)}^\infty  {\frac{{{e^{ - \lambda g}}}}{{\ln \left( {1 + \frac{{{p_T}g}}{{B{\sigma ^2}}}} \right)}}} dg{\rm{ - }}\frac{{{e^{ - \lambda G(r)}}}}{{\ln \left( {1 + \frac{{{p_T}G(r)}}{{B{\sigma ^2}}}} \right)}}
\end{equation}

For $\Xi_{(1)}(G(r))$, its partial derivative with $G(r)$ can be written as
\begin{equation}\label{partial xi1}
	\frac{{\partial {\Xi _{(1)}}(G(r))}}{{\partial G(r)}} = \frac{{{e^{ - \lambda G(r)}}\left( {\frac{{{p_T}}}{{B{\sigma ^2}}} - \lambda \ln \left( {1 + \frac{{{p_T}G(r)}}{{B{\sigma ^2}}}} \right)\left( {1 + \frac{{{p_T}G(r)}}{{B{\sigma ^2}}}} \right)} \right)}}{{{{\left( {\ln \left( {1 + \frac{{{p_T}G(r)}}{{B{\sigma ^2}}}} \right)} \right)}^2}\left( {1 + \frac{{{p_T}G(r)}}{{B{\sigma ^2}}}} \right)}}
\end{equation}
whose polarity is in coordination with the following term
\begin{equation}
	\Xi_{(2)}(G(r))=  {\frac{{{p_T}}}{{B{\sigma ^2}}} - \lambda \ln \left( {1 + \frac{{{p_T}G(r)}}{{B{\sigma ^2}}}} \right)\left( {1 + \frac{{{p_T}G(r)}}{{B{\sigma ^2}}}} \right)} 
\end{equation}
by ignoring the terms in the right-hand side of (\ref{partial xi1}) always being positive for $G(r)>0$. 
Obviously, $\Xi_{(2)}(G(r))$ is a monotonically decreasing function with $G(r)$. Moreover, it has only one zero point, which is denoted as $G^{\dagger}(r)$, since $\Xi_{(2)}(G(r))$ is monotonically decreasing, $ {\lim }_{G(r) \to 0 }\Xi_{(2)}(G(r))=p_T/(B{\sigma ^2})>0$, and ${\lim }_{G(r) \to +\infty }\Xi_{(2)}(G(r))=-\infty<0$. 
To this end, we can claim that $\frac{{\partial {\Xi _{(1)}}(G(r))}}{{\partial G(r)}}$ is positive in $(0,G^{\dagger}(r)]$ and negative in $(G^{\dagger}(r),+\infty)$. In other words, $\Xi_{(1)}(G(r))$ is increasing in $(0,G^{\dagger}(r)]$ and decreasing in $(G^{\dagger}(r),+\infty)$. 
}

Since $ {\lim }_{G(r) \to 0^+ }\Xi_{(1)}(G(r))=-\infty<0$, which can be proved by the convergence of the improper integral, and $\Xi_{(1)}(G(r))>0$ for $G(r) > x^\dagger$ as disclosed in (\ref{partial 0}), together with monotonicity of $\Xi_{(1)}(G(r))$ with $G(r)$ just explored, we can infer that $\Xi_{(1)}(G(r))$ has only one zero point, denoted as $G^\ddagger(r)$, and there is $ G^\ddagger(r) \in (0,x^\dagger]$.
% and we are sure that $ G^\ddagger(r) \in (0,x^\dagger]$. 
To this end, we can claim $\Xi_{(1)}(G(r))$ is negative for  $G(r) \in (0,G^\ddagger(r)]$ and positive for $G(r) \in (G^\ddagger(r),+\infty)$, which also implies the decreasing and increasing monotonicity of $\mathbb{E}\left\{{T_n}\right\}\mathbb{E}\left\{{P_n}\right\}$ with $G(r)$ for $G(r) \in (0,G^\ddagger(r)]$ and $(G^\ddagger(r),+\infty)$, respectively.

{
In the final step, collecting the above explored monotonicity of $\mathbb{E}\left\{{T_n}\right\}\mathbb{E}\left\{{P_n}\right\}$ with $G(r)$ when $G(r)$ varies in $(0,G^\ddagger(r)]$ or $(G^\ddagger(r),+\infty)$, recalling the increasing monotonicity of ${G(r)}$ with $r$, and defining a $r^{\ddagger}$ such that $G(r^\ddagger)=G^\ddagger(r)$, we can easily find that $\mathbb{E}\left\{{T_n}\right\}\mathbb{E}\left\{{P_n}\right\}$ is decreasing with $r$ for $r \in ( 0,r^\ddagger]$ and increasing with $r$ for $r \in (r^\ddagger, +\infty)$.
}

\bibliographystyle{IEEEtran}
\bibliography{IEEEabrv,myrefs}

\end{document}